\documentclass[12pt,preprint]{aastex}

\slugcomment{}

\shorttitle{North-South Neutrino Heating Asymmetry}
\shortauthors{Kei Kotake et al.}

\begin{document}


\title{North-South Neutrino Heating Asymmetry in Strongly Magnetized and
Rotating Stellar Cores}    

\author{Kei Kotake\altaffilmark{1}, Shoichi Yamada\altaffilmark{2},
and Katsuhiko Sato\altaffilmark{1,3}}
\affil{$^1$Department of Physics, School of Science, University of Tokyo,
7-3-1 Hongo, Bunkyo, Tokyo 113-0033, Japan}
\email{kkotake@utap.phys.s.u-tokyo.ac.jp}
\affil{$^2$Science \& Engineering, Waseda University, 3-4-1 Okubo, Shinjyuku,
Tokyo, 169-8555, Japan}
\affil{$^3$Research Center for the Early Universe, University of Tokyo, 
7-3-1 Hongo, Bunkyo, Tokyo 113-0033, Japan}



\begin{abstract}
We perform a series of two-dimensional magnetohydrodynamic simulations of supernova
cores. Since the distributions of the angular momentum and the magnetic fields of
strongly magnetized stars are quite uncertain, 
we systematically change the combinations of the strength of
the angular momentum, the rotations law, the degree of differential
rotation, and the profiles of the magnetic fields to construct the
 initial conditions. By so doing, we estimate how the
 rotation-induced anisotropic neutrino heating are affected by the
 strong magnetic fields through parity-violating effects and first investigate
 how the north-south asymmetry of the neutrino heating in a strongly magnetized
 supernova core could be. As for the microphysics, we employ a realistic 
equation of state based on the relativistic mean field theory and take
 into account electron captures and the neutrino transport via the
 neutrino leakage scheme. 
With these computations, we find that 
the parity-violating corrections reduce 
$ \lesssim 0.5 \%$ of the neutrino heating rate than that 
without the magnetic fields in the vicinity of the north pole of a star, 
on the other hand, enhance about $ \lesssim 0.5 \%$ in the vicinity of the 
south pole. If the global asymmetry of the neutrino heating in the
 both of the poles develops in the later phases, the
newly born neutron star might be kicked toward the north pole in the subsequent time.
\end{abstract}






\section{Introduction}
Neutron stars are created in the aftermath of the gravitational core
collapse 
of massive stars at the end of their lives.
After the supernova explosions, the nascent neutron
stars receive the rapid rotations and large magnetic fields, which are
believed to be observed as pulsars. 
Thus, rotations and magnetic fields should
be naturally taken into account in order not only to clarify the explosion
mechanism, but also to explain the observed properties of neutron stars.
Here it is noted that the number of studies including such multidimensional
aspects is increasing recently \citep{akiyama,buras,kotake,kotakemhd,Ott,kotakegw,yama03,kotakegw04,fryer,muller03,yamakotake,liepen}. 
This may reflect the fact that only the neutrino heating
mechanism assuming spherical symmetry seems difficult to produce the
explosions \citep{ramp,Lieben01,tomp, Lieben03}.
  
Among the unresolved mysteries in context of core-collapse supernovae, the physical origin of the
pulsar kicks has been long controversial. 
Recent analyses of individual pulsar motions and 
observations of remnant associations between supernovae and pulsars 
indicate that supernovae explosions are asymmetric and 
that the neutron stars receive large kick velocities at birth 
(typically $300 -400$ km/s \citep{lyne,lorimer}, with highest values greater than $1000$ km/s 
\citep{arzoumanian}). The existence of pulsar kicks are also
supported by the evolutionary studies of neutron star and black hole 
binaries \citep{fryerkalogera,wex,mirabel} and by the
detections of geodetic and orbital plane precessions in
some binary pulsars \citep{cordes,kapsi}.
The recent X-ray observations have shown the correlation between the 
direction of pulsar motions and the spin axis of their supernovae in Vela and
Crab pulsars \citep{helfand,pavlov}. However it is statistically uncertain whether the
spin-kick alignment is a generic feature of all pulsars. 

So far two major classes of mechanism for these kicks have been
suggested (see \citet{lai2,lai1} for a review).
Any models or mechanisms should explain how the explosions become
asymmetric and how the generated
asymmetric explosions are related to the formation of the pulsar kicks. 
The first class of the mechanisms relies on the asymmetric explosion  
as a result of the global convective instabilities
\citep{burohey,goldreich} caused by the pre-collapse density
inhomogeneities \citep{bazan}
or the local convective instabilities \citep{jankamueller94} 
formed after the onset of core-collapse. 
More recently, \citet{scheck} reported in their two dimensional simulations 
that in the long duration explosions (more than a second after bounce), the
neutrino-driven convections behind the expanding shock can lead the 
global asymmetries, which accelerates the remnant neutron star to a 
several hundreds of km/s 
(see, also \citet{fryerkick} for three dimensional calculations).
 In their simulations, the central protoneutron star is fixed 
as the inner boundary condition. Fully self-consistent calculations are 
required to confirm whether this assumption is valid or not. 

Here we pay attention to the second class of models
in which the pulsar kicks arise from an asymmetry of neutrinos 
induced by the strong magnetic field ($B \ge 10^{15}$ G) in the
collapsed supernova core. While the dipolar magnetic fields of most
radio pulsars are observed to lie in the range of $10^{12}\sim
10^{13}$ G,  
several
recent observations imply that the soft gamma-ray repeaters and
anomalous X-ray pulsars have the strong magnetic fields as high as
$B \sim 10^{15}$ G \citep{zhang,gusei}. 
So far, much theoretical efforts have been paid to formulate the
neutrino opacity in such a strong magnetic fields 
(\citet{horowitz,arras1}, \citet{arras2}). 
On the other hand, it seems that little attention has been paid to
investigate how the asymmetric neutrino heatings work globally in the
supernova core based on the numerical simulations. In addition, 
most of the preceding studies relies on 
the asymmetry of the neutrinos only by the magnetic fields (see, however,
\citet{janka99,duan04} for the simple models of atmosphere 
in the supernova core).
It is noted that rotation should also 
contribute to produce the asymmetry of 
the neutrino heating \citep{kotake}. To our knowledge,
the degree of 
the neutrino heating asymmetry under the presence of 
the rotation and the strong magnetic fields at the same time 
has not been investigated so far. 
This situation leads us to investigate how the rotation-induced anisotropic 
neutrino heating \citep{kotake} are affected by the strong magnetic
fields through parity-violating effects.
 In this study, we perform a series of two-dimensional 
simulations of magnetorotational core collapse.  
In order to see the parity-violating effects, we pay particular 
attention to the models, which are assumed to have 
the strong poloidal magnetic fields ($\sim 10^{12} $ G) 
prior to core-collapse.
Since the profiles of the angular momentum and the magnetic fields of
such strongly magnetized stars are still quite uncertain (see, however, \citet{heger03}), 
we change the strength of the rotation and the magnetic fields systematically.
Based on the 
simulations from the onset of the core-collapse through the bounce to
the shock-stall, we demonstrate how large the anisotropy of neutrino 
heating could be globally in the supernova core. 
Furthermore, we hope to speculate their possible effects on the pulsar kicks.

We describe the numerical methods in section \ref{s2}. The main
numerical results are shown in section \ref{s3}.
Discussion are given in section \ref{s4}. 
We conclude this paper in section \ref{s5}.

\section{Numerical Methods and Initial Models \label{s2}}

The numerical method for magnetohydrodynamic (MHD) computations employed
in this paper is based on the ZEUS-2D code \citep{stone}. The ZEUS-2D is an
Eulerian code based on the finite-difference method and employs an
artificial viscosity of von Neumann and Richtmyer to capture shocks.
The time evolution of magnetic field is solved by the
induction equation : $\partial  \mbox{\boldmath$B$}/\partial t = 
\nabla \times( \mbox{\boldmath$v$}\times \mbox{\boldmath$B$})$, with  
$\mbox{\boldmath$v$}, \mbox{\boldmath$B$}$ being velocity and the
magnetic field, respectively. In so doing, the code utilizes 
the so-called constrained transport (CT) method, which ensures 
that the numerically evolved magnetic fields are  
divergence-free ($\nabla \cdot \mbox{\boldmath$B$} = 0$) at all times. Furthermore, the method of characteristics (MOC) is
implemented to propagate accurately all modes of MHD waves.
The self-gravity is managed by solving the Poisson equation with the
incomplete Cholesky decomposition conjugate gradient 
method. In all the computations, spherical coordinates are used and 
one quadrant of the meridian
section is covered with 300 ($r$) $\times$ 50 ($\theta$) mesh points.
We have made several major changes to the base code to incorporate the 
microphysics. First, we added an equation for the electron fraction to
treat electron captures and neutrino transport by the so-called leakage
scheme \citep{ep,blud,van1,van2}. The calculation of electron fraction
is done separately from the main hydrodynamic step in an
operator-splitting manner. Second, we implemented a tabulated
equation of state based on the relativistic mean field theory \citep{shen98}
instead of the ideal gas EOS assumed in the original code. For a more
detailed description of the methods, see \citet{kotake}.
\subsection{Initial Conditions}
Recently, \citet{heger03} performed the most up-to-date stellar evolution 
calculations including the effects of rotation, mixing, transport of the
angular momentum, and the magnetic torques. Their models
show that the toroidal magnetic fields are much stronger than the
poloidal ones prior to core collapse. In our previous paper 
\citep{kotakemhd}, we considered such models as the initial conditions.
However, if such initial configurations are true, 
no parity-violating effects of the magnetic fields on the neutrino
heating can be obtained in the two-dimensional
axisymmetric simulations. Considering the uncertainties in Heger's evolution
calculations that they are one-dimensional models, which hinder the
accurate treatment of the transport of the angular momentum and the generation
of the magnetic fields, it should be still important to 
take the poloidal magnetic fields as the initial conditions. 
Thus in this study, we choose to assume that only the poloidal magnetic
fields exist prior to core-collapse, and then, prefer a parametric
approach to construct the initial conditions varying the profiles of the
initial poloidal magnetic fields. 
We assume the following two rotation laws and the four configurations
of the magnetic fields. 

1. shell-type rotation:
\begin{equation}
\omega(r) = \omega_0 \times \frac{{R_{0}}^2}{r^2 + R_{0}^2},
\end{equation}
   and shell-type magnetic fields,
\begin{equation}
B(r) = B_0 \times \frac{{R_{0}}^2}{r^2 + R_{0}^2},
\end{equation}
where $\omega(r)$ is an angular velocity, $B(r)$ is the poloidal
component of the magnetic fields, $r$ is radius, and
$\omega_0, R_{0}$ and $B_{0}$ are model constants,  

2. cylindrical rotation:
\begin{equation}
\omega(X,Z) = \omega_0 \times \frac{{X_{0}}^2}{X^2 + {X_{0}}^2} \cdot
\frac{Z_{0}^4}{Z^4 + Z_{0}^4},
\end{equation}
   and cylindrical magnetic fields,
\begin{equation}
B_{z}(X,Z) = B_0 \times \frac{{X_{0}}^2}{X^2 + {X_{0}}^2} \cdot
\frac{Z_{0}^4}{Z^4 + Z_{0}^4},
\end{equation}
  and uniform cylindrical magnetic fields,
\begin{equation}
B_{z} = B_{0},
\end{equation}
where $B_z$ is the poloidal magnetic fields parallel to the rotational
($Z$) axis, $X$ and $Z$ denote distances from the rotational axis and the equatorial plane and $X_0, Z_0$ are model constants. The other parameters have the same meanings as above.
 
Finally, we take into account the quadrupole magnetic fields following
the prescription by \citet{ardeljan1},
\begin{eqnarray}
B_x(X,Z) &=& F_x(0.5\hat{X}, 0.5\hat{Z}-2.5) - F_x(0.5\hat{X}, 0.5\hat{Z}+2.5), \nonumber\\
B_z(X,Z) &=& F_z(0.5\hat{X}, 0.5\hat{Z}-2.5) - F_z(0.5\hat{X}, 0.5\hat{Z}+2.5),
\end{eqnarray}
with
\begin{equation}
\hat{X} = \frac{X}{X_{0, {\rm quad}}},~~\hat{Z} = \frac{Z}{Z_{0, {\rm quad}} },
\end{equation}
and
\begin{eqnarray}
F_x(\hat{X},\hat{Z}) &=& 
B_0\Biggl(\frac{2\hat{X}\hat{Z}}{(\hat{Z}^2+1)^3} - \frac{2\hat{X}^3\hat{Z}}{(\hat{Z}^2+1)^5}\Biggr) \nonumber \\
F_z(\hat{X},\hat{Z}) &=& 
B_0\Biggl(\frac{1}{(\hat{Z}^2+1)^2} - \frac{\hat{X}^2}{(\hat{Z}^2+1)^4}\Biggr),
\end{eqnarray}
where $B_{x}$ is the poloidal magnetic fields parallel to the equatorial
  plane and $X_{0, {\rm quad}},Z_{0, {\rm quad}}$ are model constants.

  We have computed 11 models changing the combination of the strength of
  the angular momentum, the rotation
  law, the degree of differential rotation, and the profiles of the
  magnetic fields. For all the models, we fix the values of
  $E_{\rm m}/|W|$ to be $0.1\%$, where $E_{\rm m}/|W|$ represents the
  ratio of the magnetic to the gravitational energy.
 The model parameters are presented in Table
 \ref{table:1}. The models are named after this combination, with the
 first letter ,``S (shell)'', ``C (cylindrical)'', denoting the rotation
  profiles, the second letter ,``L
 (long), S (short)'', indicating the values of $R_{0}$ and $X_{0}$, the
  third letter ``(U) uniform, (NU) non uniform, (Q) quadrupole''
  indicating the configurations of the initial poloidal
  magnetic field, the fourth number -1 or -2, representing that the initial
  $T/|W|$ is $10^{-1}, 10^{-2} \%$, respectively. Note that the ratio of rotational energy to gravitational 
  energy is designated as $T/|W|$.
  
We note that our choice of the initial value of $T/|W|$ is much smaller
than the past magnetorotational studies in context of core-collapse
supernovae \citep{leblanc,Bisno,muller,sym,ard} to reconcile with the recent 
stellar evolution calculations of rotating stars \citep{heger00}. 
However it is still larger than the most slowest rotating models predicted 
by the most recent stellar calculations \citep{heger03}. 
As for the
configurations of the initial magnetic fields,  we 
prepared a variety of them in order to cover as wide the parametric
space of the field configurations as possible. 
As for the initial strength of the magnetic fields, 
we pay particular attention to the models, which are assumed to have 
the strong poloidal magnetic fields ($\sim 10^{12} $ G) 
prior to core-collapse in order to see the parity-violating effects.
Astrophysically, such models are likely to associate with the formation of the so-called magnetars.

\begin{deluxetable}{lccccccc}
\tabletypesize{\scriptsize}
\tablecaption{The Model Parameters. \label{table:1}}
\tablewidth{0pt}
\tablehead{
\colhead{Model} & \colhead{Rotation profiles}  & \colhead{B profiles}   
 & \colhead{$\omega_0\,\,(\rm{s}^{-1})$}   &
\colhead{$B_0\,\,(\rm{G})$} &
\colhead{ $ R_{0}$, $X_{0}$, $Z_0$ $\times 10^8$ (cm)}  & \colhead{  }}
\startdata
 SL(U)-2 & Shell    & Uniform Cylinder
          &  0.6          & $1.9 \times 10^{12}$   & $R_{0} = 1$  &  \\
 SS(U)-2        & Shell    & Uniform Cylinder
          &  9.0          & $1.9 \times 10^{12}$   &  $R_{0} = 0.1$ &  \\
 CS(U)-2        & Cylinder & Uniform Cylinder         
          &  6.3      &   $1.9 \times 10^{12}$  & $X_{0} = 0.1,Z_{0}=1$  & \\
 SS(NU)-2      & Shell    & Shell 
          &  9.0          & $1.1 \times 10^{14}$  &  $R_{0} = 0.1$  & \\
 CS(NU)-2       & Cylinder & Cylinder
          &  6.3      &  $4.7 \times 10^{13}$     &  $X_{0} = 0.1,Z_{0}=1$  &  \\
 SL(Q)-2        & Shell    & Quadrupole
          &  9.0  & $1.4\times 10^{13}$    & $R_{0} = 1$, 
$X_{0,{\rm quad}}=Z_{0,{\rm quad}} = 1.6 $ &  \\
 CS(Q)-2        & Cylinder & Quadrupole
          &  6.3    &  $1.4\times 10^{13}$    & $X_{0} = 0.1,Z_{0}=1$, $X_{0,{\rm quad}}=Z_{0,{\rm quad}} = 1.6 $&  \\
 SL(U)-1 & Shell    & Uniform Cylinder
          &   1.7         & $1.9 \times 10^{12}$   & $R_{0} = 1$  & \\
 SS(U)-1        & Shell    & Uniform Cylinder
          &   28.3         &  $1.9 \times 10^{12}$  & $R_{0} = 0.1$ &  \\
 SL(NU)-1& Shell    & Shell    
          &   1.7     &  $5.2 \times 10^{12}$   & $R_{0} = 1$   &  \\
 SS(NU)-1      & Shell    & Shell 
          &   28.3    & $1.1 \times 10^{14}$  & $R_{0} = 0.1$  &  \\

\enddata
\tablecomments{Note in the table that ``B profiles'' indicates the
 profiles of the initial magnetic fields. Note also that the number
(-1 or -2), which each model has in its name, indicates the values of
 initial  $T/|W|$ of $10^{-1}, 10^{-2} \%$, respectively.}
\end{deluxetable}

\begin{deluxetable}{lccccccc}
\tabletypesize{\scriptsize}
\tablecaption{Properties of the Final States.\label{table2} }
\tablewidth{0pt}
\tablehead{
\colhead{Model} & \colhead{$t_{\rm f}$ (ms)}& \colhead{$B_{p}$ (G)}
& \colhead{$B_{\phi}$ (G)}   &
 \colhead{$r_{\rm{sh}}^{\rm{e}}$ (km)}   
& \colhead{$ r_{\rm{sh}}^{\rm{p}}$ (km)}  
& \colhead{$T/|W|_{\rm f} (\%)$}   
& \colhead{$E_{\rm m}/|W|_{\rm f} (\%)$}}
\startdata
SL(U)-2 &  40.2   & $1.5\times 10^{16}$  & $2.4\times 10^{15}$       
 & 157 & 175 & $2.5 \times 10^{-1}$ & $1.9 \times 10^{-1}$\\
SS(U)-2 &  41.8   & $3.4 \times 10^{16}$  & $2.3 \times 10^{15}$      
 & 178 & 233  & $2.5\times 10^{-1}$ & $1.7 \times 10^{-1}$ \\
CS(U)-2 &  40.7   &  $3.2 \times 10^{16}$ & $3.7 \times 10^{15}$            
 & 144 & 612  & $2.5 \times 10^{-1}$ & $1.7 \times 10^{-1}$  \\
SS(NU)-2 & 42.5   & $5.2 \times 10^{16}$  & $1.9 \times 10^{15}$           
 & 163 & 148 & $1.1 \times 10^{-1}$ & $5.3 \times 10^{-1}$\ \\
CS(NU)-2 & 43.1   & $4.6 \times 10^{16}$  & $4.9 \times 10^{14}$ 
 & 176 & 1471 & $1.1 \times 10^{-1}$ & $3.7 \times 10^{-1}$\\
SL(Q)-2 & 37.1    & $1.2 \times 10^{16}$  & $9.4 \times 10^{15}$          
 & 146 & 125  &$2.6\times 10^{-1}$ & $5.2\times 10^{-2}$\\
CS(Q)-2 & 46.0    & $3.0\times 10^{16}$     & $9.3\times 10^{15}$            
 & 169 & 215 & $3.4\times 10^{-1}$ & $7.1 \times 10^{-2}$\\
SL(U)-1 & 42.3       &  $3.0 \times 10^{16}$ & $6.0\times 10^{15}$            
 & 226 & 573 & 1.4  & $3.8 \times 10^{-1}$ \\
SS(U)-1 & 47.1        &  $4.1 \times 10^{16}$ & $2.0 \times 10^{16}$          
&  133 & 1361  & 1.8 &  $3.6 \times 10^{-1}$  \\
SL(NU)-1 & 45.0  & $2.7 \times 10^{16}$      &$4.9 \times 10^{15}$     
&  278 & 444 & 1.0   & $8.3 \times 10^{-1}$ \\
SS(NU)-1 & 44.4        & $5.8\times 10^{16}$  & $7.0\times 10^{14}$    
&  425 & 1325 & $2.8\times 10^{-1}$ & $8.6 \times 10^{-1}$ \\

\enddata



\tablecomments{ $ t_{\rm f} $ represents the final time measured from core bounce. 
$B_{p}$ and $B_{\phi}$ are the maximum strength of the poloidal
 and the toroidal magnetic fields, respectively.
$r_{\rm{sh}}^{\rm{e}}$ and 
$r_{\rm{sh}}^{\rm{p}}$ are the distances from the center to the stalled
 shock front in the equatorial plane and along the rotational axis,
 respectively.}

\end{deluxetable}

\begin{deluxetable}{lccccc}
\tabletypesize{\scriptsize}
\tablecaption{Analysis of the Heating Rate 
in the Strong Magnetic Fields.
\label{heat_hikaku} }
\tablewidth{0pt}
\tablehead{
\colhead{Model} & \colhead{${R_{\nu}}^{\rm{p}}$ (km)} &   
 \colhead{${R_{\nu}}^{\rm{e}}$ (km)}       &
 \colhead{ $R_{\rm mag}~(\%)$}}
\startdata
SL(U)-2    & 66.6  & 73.1  & 0.13   \\
SS(U)-2    & 71.4  & 71.4  & 0.27   \\
CS(U)-2    & 57.3  & 73.1  & 0.27   \\
SS(NU)-2   & 65.1  & 68.2  & 0.08   \\
CS(NU)-2   & 54.3  & 68.2  & 0.18   \\
SL(Q)-2    & 69.8  & 68.2  & 0.14   \\
CS(Q)-2    & 66.6  & 68.2  & 0.09   \\
SL(U)-1    & 57.3  & 73.1  & 0.37   \\
SS(U)-1    & 55.8  & 66.6  & 0.42   \\
SL(NU)-1   & 50.0  & 65.1  & 0.26    \\
SS(NU)-1   & 51.4  & 66.6  & 0.12    \\
\enddata           



\tablecomments{${R_{\nu}}^{\rm{e}}$ and 
${R_{\nu}}^{\rm{p}}$ are the distances from the center to the neutrino
 sphere, respectively. $ R_{\rm mag} = |\Delta Q_{\nu, \rm{B} \neq 0}^{+}/Q_{\nu, \rm{B} = 0}^{+}|$ is the maximum ratio of the neutrino heating rate contributed from the parity-violating effects, $\Delta Q_{\nu, \rm{B} \neq 0}^{+}$ to the heating rate without the corrections, $Q_{\nu, \rm{B} = 0}^{+}$.}

\end{deluxetable}

\begin{figure}
\epsscale{1.0}
\plotone{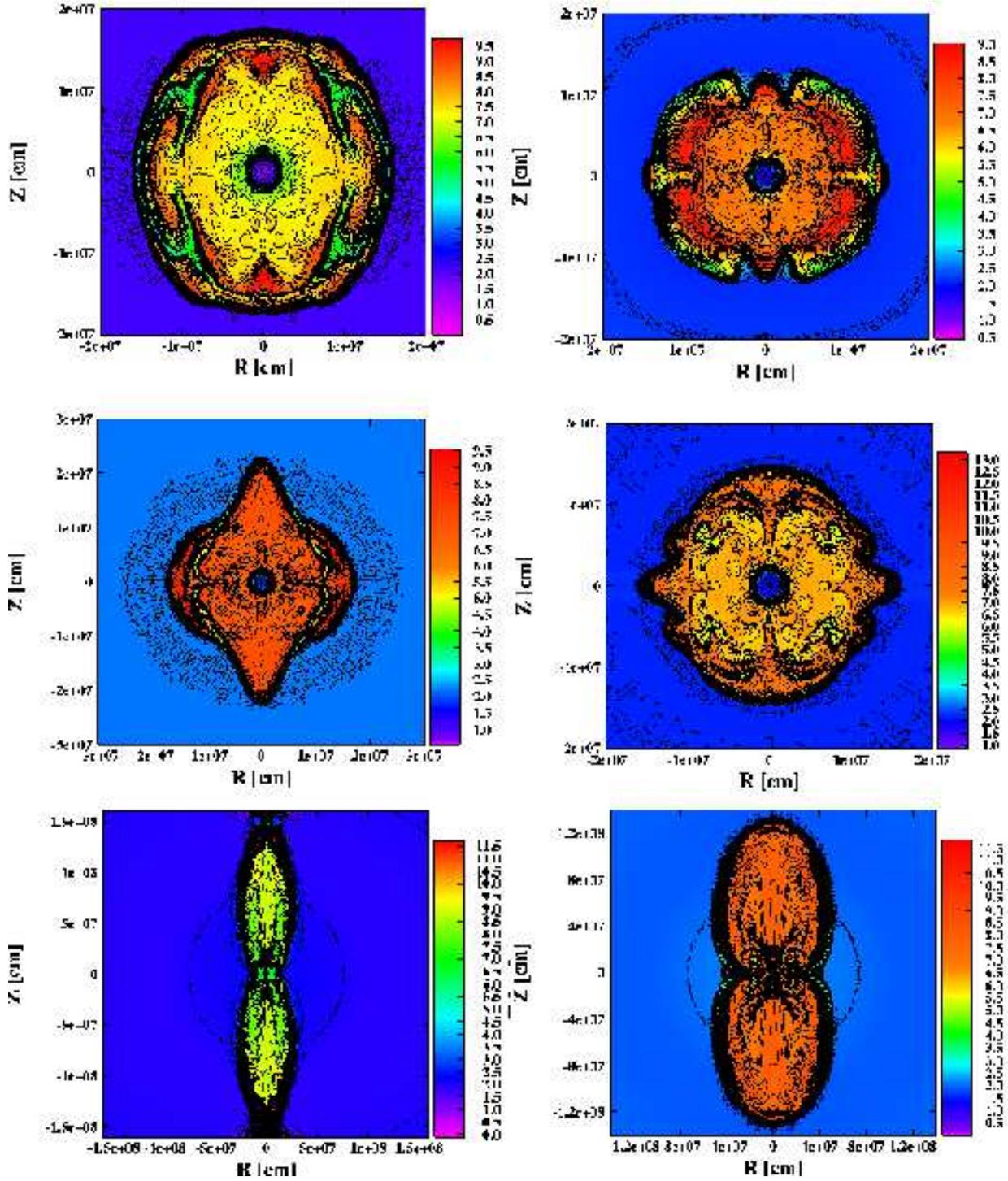}
\caption{The final profiles of models SL(U)-2 (upper left), SS(Q)-2 (upper
 right), SS(U)-2 (middle left), SS(NU)-2 (middle right), CS(NU)-2 (lower
 left), SS(NU)-1 (lower right). They show color-coded contour
 plots of entropy ($\rm{k_{B}}$) per nucleon.}
\label{fig1}
\end{figure}

\begin{figure}
\begin{center}
\plotone{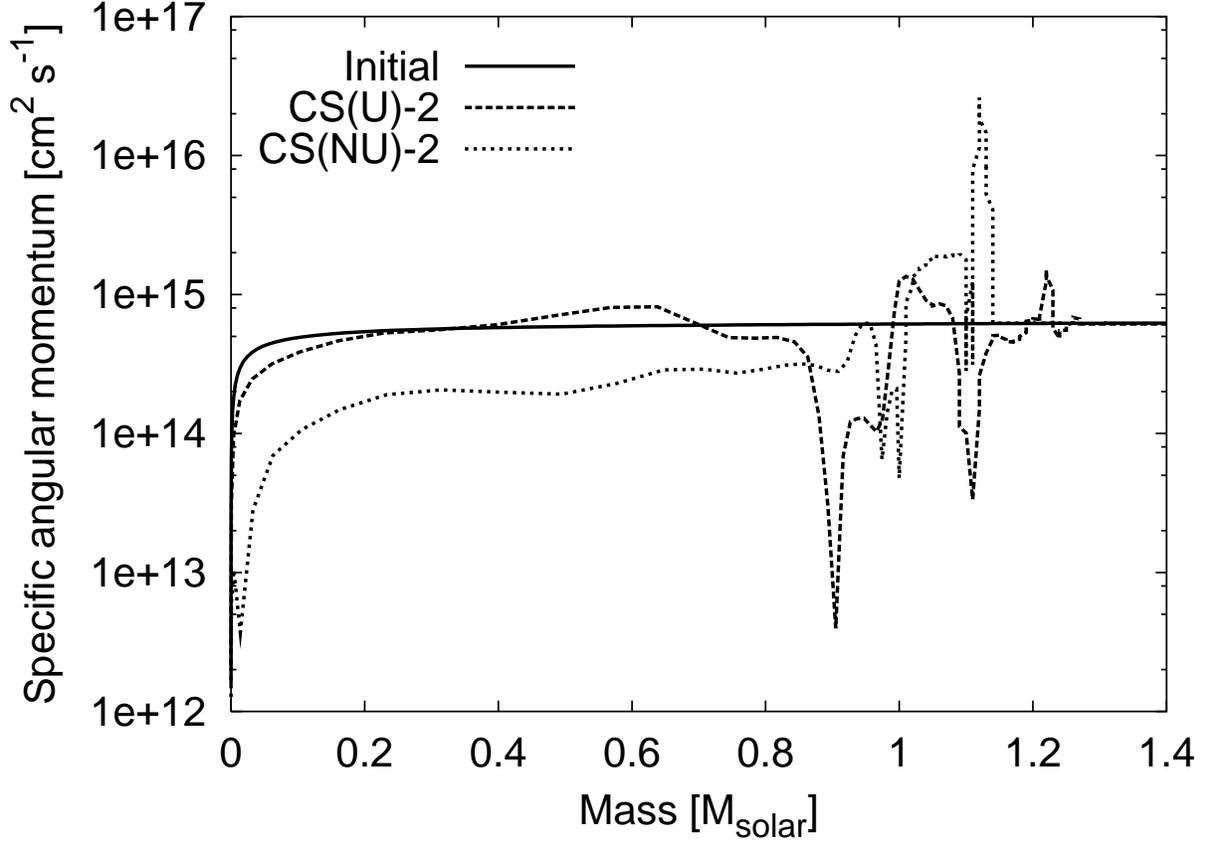}
\caption{The cumulative mass vs. specific angular momentum distribution
 in unit of $M_{\odot}$
 for models CS(U)-2 and CS(NU)-2 at the final state. 
Note in the figure that 'Initial'
 represents the initial configuration of the specific angular
 momentum for the models. It is seen that the angular momentum is transferred more
 efficiently for the model with the centrally condensed magnetic fields
 profile (model CS(NU)-2) than model CS(U)-2.}
\label{fig2}
\end{center}
\end{figure}

\begin{figure}
\begin{center}
\plottwo{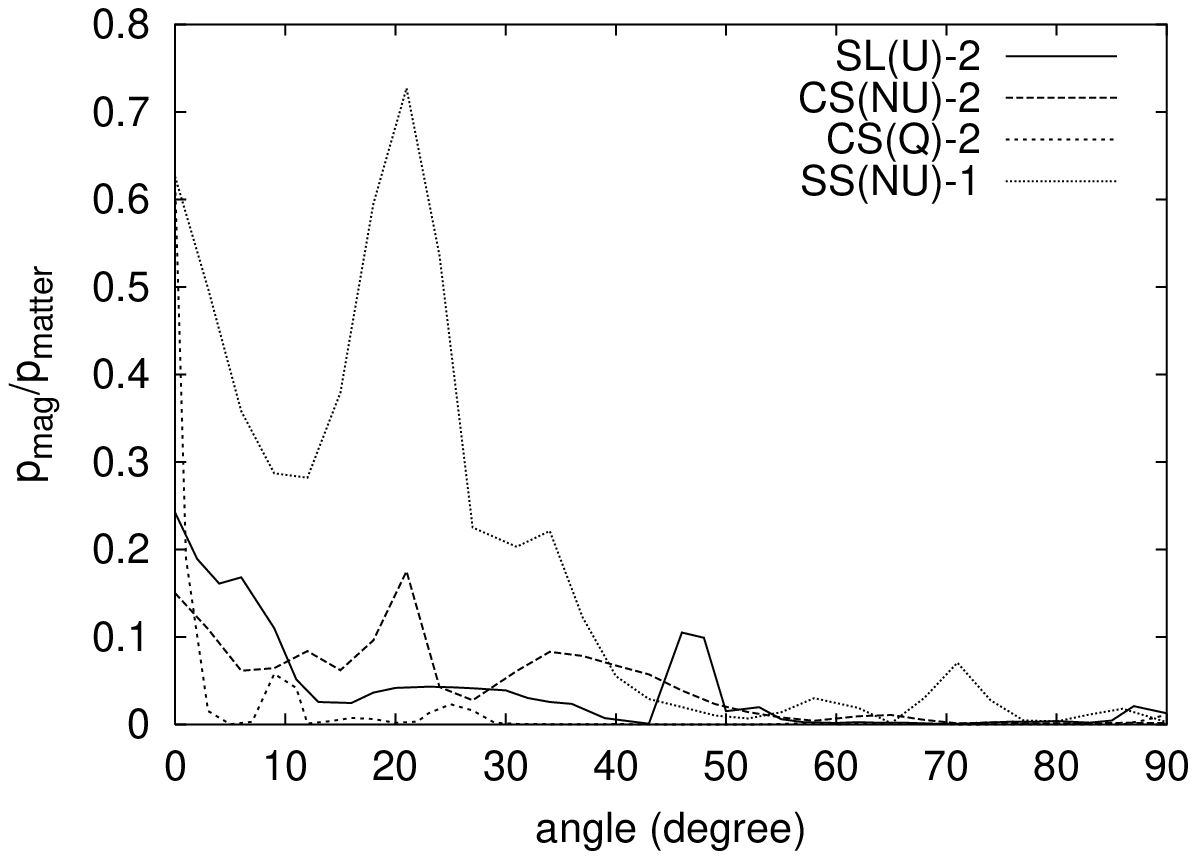}{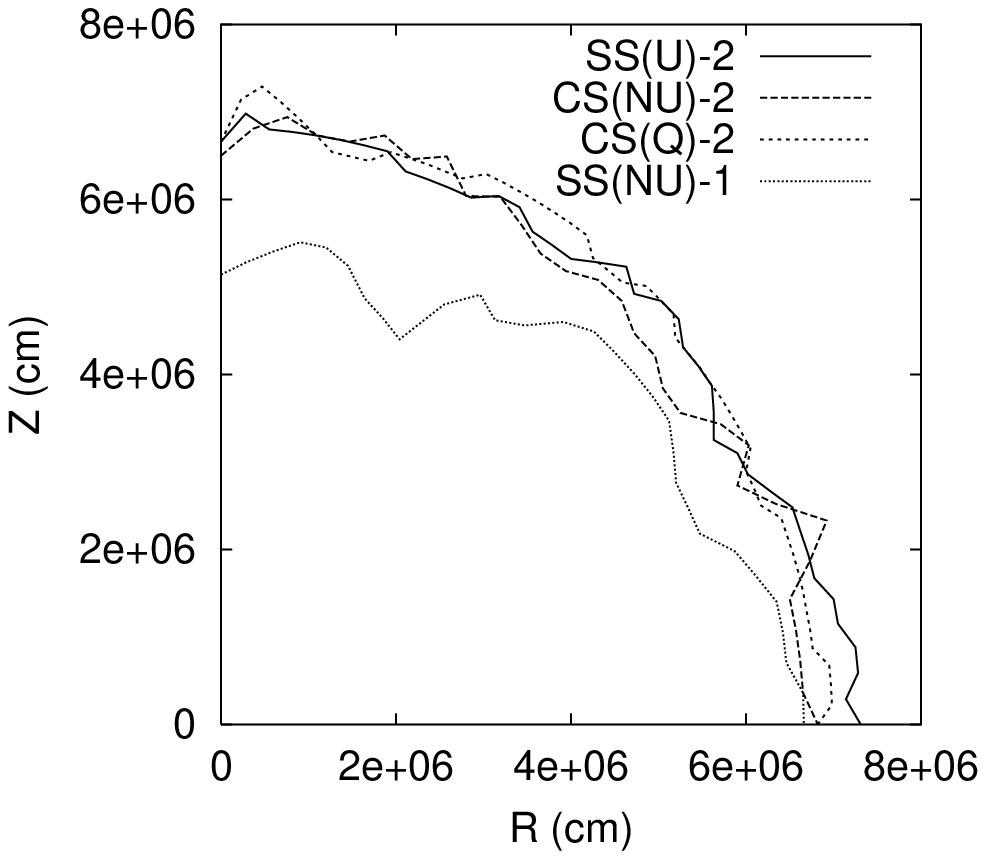}
\caption{The ratio of the magnetic ($p_{\rm mag}$) to the matter pressure
 ($p_{\rm matter}$) with the polar angle on the neutrino sphere (left
 panel) and the shapes of the neutrino spheres for some representative
 models (right panel).  Note in the right panel that $R, Z$ represents the distance
 from the rotational axis and the equatorial plane, respectively.}
\label{fig3}
\end{center}
\end{figure}

\section{Results \label{s3}}
\subsection{Magnetohydrodynamic Features}
We briefly summarize some dynamical aspects which will be helpful in the
later discussions. For the more detailed magnetohydrodynamic computations
during core-collapse,
see also our accompanying paper \citep{takiwakietal}.

In Figure \ref{fig1}, the entropy distributions at the final states for some
representative models are presented.
The final time of all the models is
about 50 ms after core bounce (see Table \ref{table:1}).
At the final time, the shock wave produced at core bounce tends to stall
in the core except for the models, which have the centrally condensed
magnetic fields and rapid rotation with the strong differential rotation initially (see Figure \ref{fig1} and Table
\ref{table:1} for models CS(NU)-2, SS(U)-1, and SS(NU)-1).  Even if the shock waves of such models 
manage to approach the outer boundary of the numerical simulation 
($Z \sim 1500$ km) along the rotational axis, the explosion energy is about one order of magnitude 
less than the canonical one. Thus, the neutrino heatings in the later
phases should be necessary for the successful explosions.
Here it is noted that the final value of $T/|W|$'s and $E_{\rm
m}/|W|$'s for our models are well below the past MHD studies \citep{sym}
assuming rapid rotation prior to core-collapse. Even 
in our slower rotating models than the previous simulations, 
it is found that the asymmetric matter motions can appear as seen from
Figure \ref{fig1}. 
  
During core-collapse, the strength of the magnetic fields 
exceeds the QED critical value $B_{\rm QED} = 4.4 \times 10^{13}$ (G),
above which the neutrino reactions are affected by
the parity-violating corrections of weak interaction.
Just before core bounce, the poloidal 
magnetic fields dominate over the toroidal ones due to
our assumption of the initial field configuration.
After core bounce, the strong differential rotations near the surface of
the protoneutron star wraps the poloidal magnetic fields 
to convert into the toroidal ones, whose strength becomes 
as high as $ B \sim 10^{15}$ G at the final time of the simulations.
The growth of the toroidal magnetic fields can be estimated by an
order-of-magnitude estimation as follows, 
\begin{equation}
B_{\phi} \sim 2.0\times 10^{15}~{\rm G}~\Bigl(\frac{B_{z}}{10^{15}~{\rm G}}\Bigr)\Bigl( ~\frac{\Delta t}{50~{\rm msec}}\Bigr)\Bigl( 
~\frac{\omega}{500 ~{\rm rad}~\rm{s}^{-1}}\Bigr),
\end{equation} 
where $B_{z}$ is the typical strength of the magnetic field parallel
to the rotational axis just before core bounce, $\Delta t$ is the
duration of the wrapping time, which is roughly 
equivalent to the time from core bounce to the final time, $\omega$ is
the mean angular velocity at the surface of the protoneutron star.
This estimation is roughly consistent with the numerical simulations.
It is noted that the tangled toroidal magnetic fields  
($B_{\phi} \lesssim 10^{16}~\rm{G}$)
are weaker than the poloidal ones ($B_{ p} \gtrsim 10^{16}~\rm{G} $) at the
final time in our models (see Table \ref{table:1}).
Magnetorotational instability (MRI) is an another candidate to amplify
the growth of the toroidal magnetic fields. Although the typical time
scale \citep{balbus1} corresponding to the maximum growth of MRI ($\tau_{\rm MRI} \sim
O(10)~\rm{ms} $) at the surface of the
protoneutron star is shorter than the
simulation time scale ($\sim 50$ ms), it is much longer than 
the growth time scale of the field wrapping ($\tau_{\rm wrap} \sim
O(1)~{\rm ms}$).
Thus the amplification of the toroidal magnetic fields in our models 
is mainly as a result of the winding up original vertical poloidal fields. 

In Figure \ref{fig2}, the cumulative distribution of specific
angular momentum is shown for models CS(U)-2 and CS(NU)-2. In the
magnetorotational core-collapse, the anisotropic magnetic stress
transfers the specific angular momentum of each fluid element outwards
even in axisymmetry if the fluid rotates differentially.  It is noted
that the mass of the unshocked core is about $\sim 0.8 M_{\odot}$ 
in these models. From the figure, it is
found that the transport of the angular momentum occurs more efficiently 
for the model CS(NU)-2, whose initial magnetic fields are more centrally
condensed, than the model CS(U)-2 (compare the specific angular momentum
in Figure \ref{fig2}). 
In addition, there are regions just
outside the unshocked core where
the specific angular momentum decreases outwards. 
In such regions, the so-called Rayleigh instability condition
holds, which induces the convection at the later phases near the
surface of the unshocked core in the equatorial plane. 
These differences mentioned above stem only from 
the degree of the initial concentration of the magnetic fields between
the pair models and are 
common to the other pair of the models.

After the final time of our simulations, we calculate the shapes of the neutrino spheres
by the prescription \citep{kotake}.  Although the magnetic fields
become strong 
as $B \sim 10^{15}$ G
in the vicinity of the neutrino sphere ($r = 50 \sim 70$ km, density $
\sim 10^{11} ~\rm{g}~\rm{cm}^{-3}$), the
pressure of the magnetic fields ($p_{\rm mag} \equiv \frac{B^2}{8 \pi}
\sim 10^{29} (B/{2\times 10^{15}}~\rm{G})^2~\rm{dyn}~\rm{cm}^{-2}$) are
smaller than that of the matter ($p_{\rm matter} \sim
10^{30}~\rm{dyn}~\rm{cm}^{-2}$) in the corresponding region (see the
left panel of Figure \ref{fig3}). Thus it is found that the magnetic
fields do not affect the shape of the neutrino sphere significantly.
In fact, the neutrino spheres are deformed to be oblate for the rapidly
rotating models, on the other hand, to be rather
spherical for the slowly rotating models (see the right panel of Figure
\ref{fig3} and compare the shape of the neutrino sphere of model
SS(NU)-1 (:faster rotating model in our computation) with that of 
the other models (:slower rotating models)).

\subsection{Effects of the strong magnetic fields on the neutrino
  heating}
Based on the neutrino spheres obtained in the previous section, we will
estimate the heating rates including the corrections from the
parity-violating effects outside the neutrino sphere. This is
admittedly a very crude estimate, since it is pointed out
\citep{Lieben01,tomp} that the net heating, the heating minus cooling,
becomes positive $ \sim 50 - 100$ msec after the shock stagnation. In
this respect, we confirmed that the cooling dominates over the heating
also for our models including the corrections of the 
parity-violating effects.
Since our calculations ended before the heating dominates over
the cooling, it is impossible to estimate the location of the
gain radius, beyond where the heating dominates over the
cooling. Here it is noted that the contrast of the heating to the
cooling is found to be similar in our short simulation runs. 
Thus, the gain radius should be also deformed in the later phase. 
If this feature lasts in the later phase, the bare heating rate 
discussed in the following and the resultant neutrino asymmetry due to 
the parity-violating corrections, 
will help to understand the net neutrino asymmetry.
Hence, bearing this caveat in mind, we study
the bare heating rate with the parity-violating effects 
for the final configurations (several tens msec
after the bounce) in our simulations.

\begin{figure}
\plotone{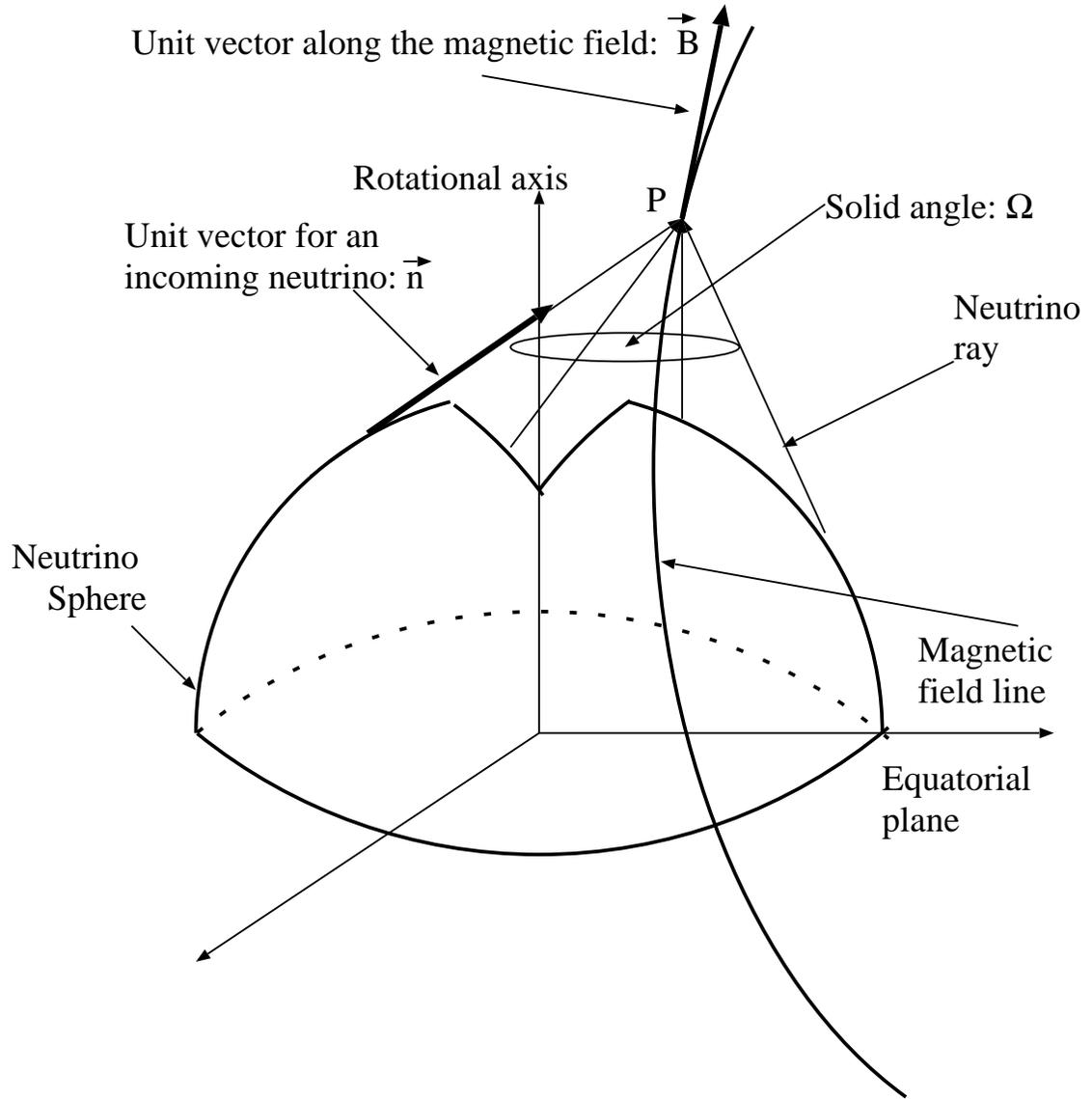}
\caption{Schematic drawing of the analysis scheme of the heating rate. Neutrinos are assumed to be emitted isotropically from each point on the neutrino sphere. The heating rate at P can be found by summing up the contributions of all rays,  reaching P.}
\label{neusp3dim}
\end{figure}

First of all, we state how we estimate the neutrino heating rate in the strongly
magnetized supernova core. The analysis scheme is schematically shown in Figure
 \ref{neusp3dim}.
 Neutrinos are assumed to be emitted isotropically from each point on
 the neutrino sphere and stream freely later on. Then the heating rate
 at a given point outside the neutrino sphere can be found by summing up
 all the rays from the neutrino sphere.
The heating by electron neutrinos proceeds via the charged current interaction:
\begin{equation}
\nu_{e} + n \rightarrow p + e^{-}.
\end{equation} 
 We employ the corresponding cross section in a strongly magnetic field
given by \citet{arras1}, which is expressed as follows : 
\begin{equation} 
{\sigma_{\rm{B}}}^{\rm (abs)} = {\sigma_{0}}^{\rm (abs)}(~1 + \epsilon_{\rm abs}~
{\mbox{\boldmath$\hat{n}$}} \cdot \mbox{\boldmath$\hat{\rm{B}}$} ),
\label{B_sigma}
\end{equation}
 where ${\sigma_{0}}^{\rm (abs)}$ is the corresponding cross section
 without the magnetic fields, ${\mbox{\boldmath$\hat{n}$}}$,
 $\mbox{\boldmath$\hat{\rm{B}}$}$ are the unit momentum vector for the
 incoming neutrinos and the unit vector along the magnetic field
 direction (see Figure \ref{neusp3dim}), and $\epsilon_{\rm abs}$ is the asymmetry parameter given by
\begin{equation}
\epsilon_{\rm abs} = \epsilon_{\rm abs}({\rm{e}}) + \epsilon_{\rm abs}({\rm {np}}).
\end{equation}

Here $\epsilon_{\rm abs}({\rm{e}})$ and $\epsilon_{\rm abs}{\rm{(np)}}$
are expressed in the leading order at the neutrino energies as 
\begin{equation}
\epsilon_{\rm abs}({\rm{e}}) = \frac{1}{2}~\frac{\hbar c~e B}{\epsilon_{\nu}^2}~\frac{c_{\rm V}^2 - c_{\rm A}^2}{{c_{\rm V}^2 + 3 c_{\rm A}^2}},
\end{equation}
and 
\begin{eqnarray}
\epsilon_{\rm abs} ({\rm np}) &=& 2~\frac{c_{\rm A}(c_{\rm A} + c_{\rm V})}
{c_{\rm V}^2 + 3 c_{\rm A}^2}~\frac{\mu_{\rm B n}  B}{k_{\rm B}~T} -
\frac{k_{\rm B}~T}{\epsilon_{\nu}}\Bigl
[ 1 + \frac{\epsilon_{\nu}}{k_{\rm B}~T}\Bigl]\nonumber \\
& & \times \Bigl[2~\frac{c_{\rm A}(c_{\rm A} + c_{\rm V})}
{c_{\rm V}^2 + 3 c_{\rm A}^2}~\frac{\mu_{\rm B n}B}{k_{\rm B} T} + 2~\frac{c_{\rm A}(c_{\rm A} - c_{\rm V})}
{c_{\rm V}^2 + 3 c_{\rm A}^2}~\frac{\mu_{\rm B p}B}{k_{\rm B} T}\Bigr],
\end{eqnarray}
which represent the
effects of electron Landau levels in magnetic fields and polarizations of
neutron and proton, respectively. Here $k_{\rm B}$ is the Boltzmann
constant, $\mu_{\rm B n} = -1.913~\mu_{\rm N}$, $\mu_{\rm B p} =
2.793~\mu_{\rm N}$ are the magnetic moments of
neutrons and protons, respectively, with $\mu_{\rm N}$ being the nuclear magneton,
$c_{\rm{V}} = 1, c_{\rm A} = 1.26$ are the coupling constants. At the considered densities and
temperatures, fermion phase space blocking and dense-medium effects can
be safely ignored. The heating rate is given as 
\begin{equation}
{Q^{+}}_{\nu} = \int {\sigma_{\rm{B}}}^{\rm (abs)}~c~n_{\rm n} \epsilon_{\nu}
\frac{d^2 n_{\nu}}{d\epsilon_{\nu} d\Omega} d\epsilon_{\nu}d\Omega,
\label{1}
\end{equation}
where $n_{\rm n}$ is the number density of neutron, $d^2{n_{\nu}}/(d\epsilon_{\nu}~d\Omega)$ is related to the
neutrino distribution in the phase space $f(\epsilon_{\nu}, \Omega)$: 
\begin{equation}
\frac{d^2 n_{\nu}}{d\epsilon_{\nu} d\Omega} = \frac{1}{(hc)^3} f_{\nu}(\epsilon_{\nu}, \Omega)~\epsilon_{\nu}^2.
\label{2}
\end{equation}
Introducing Eq.(\ref{2}) to (\ref{1}) yields
\begin{equation}
{Q^{+}}_{\nu} = \frac{c n_{n}}{(hc)^3} \int {\sigma_{\rm{B}}}^{\rm (abs)} \epsilon_{\nu}^3~f_{\nu} (\epsilon_{\nu},\Omega)~d\Omega~d\epsilon_{\nu}.
\label{heat2}
\end{equation}
Introducing Eq. (\ref{B_sigma}) to (\ref{heat2}) yields 
\begin{eqnarray}
{Q^{+}}_{\nu} &=& \frac{c n_{n}}{(hc)^3} \int {\sigma_0}^{\rm (abs)} \epsilon_{\nu}^3~f_{\nu}(\epsilon_{\nu},\Omega)~d\Omega~d\epsilon_{\nu} \nonumber\\ 
& & + \frac{c n_{n}}{(hc)^3} \int {\sigma_0}^{\rm (abs)}~\epsilon_{\rm abs}~ 
{\mbox{\boldmath$\hat{n}$}} \cdot \mbox{\boldmath$\hat{\rm{B}}$}~
\epsilon_{\nu}^3~f_{\nu}(\epsilon_{\nu},\Omega)~d\Omega~d\epsilon_{\nu} \nonumber \\
&\equiv& Q_{\nu, B = 0}^{+} + \Delta Q_{\nu, B \neq 0}^{+}.
\label{separate}
\end{eqnarray}
The first term in Eq. (\ref{separate}) can be expressed,
\begin{equation}
Q_{\nu, B = 0}^{+} = \frac{3c_{\rm A}^2 +1 }{4} \frac{\sigma_{0}\,c\,n_{\rm n}}
{(hc)^3} \frac{\langle\epsilon^2_{\nu}\rangle}{({m_{\rm e}} c^2)^2}\,\,\Omega 
\,\int \! d \epsilon_{\nu} 
\,\,\epsilon^3_{\nu}~f_{\nu}({\epsilon_{\nu}}),
\label{bef_final}
\end{equation}
where
\begin{equation}
\langle\epsilon^2_{\nu}\rangle = \int \!d\epsilon_{\nu} \,\epsilon^5_{\nu} \,f_{\nu}(\epsilon_{\nu}) \Bigl( \int \!d\epsilon_{\nu} \, \epsilon^3_{\nu} f_{\nu}(\epsilon_{\nu}) \Bigr)^{-1}.
\end{equation}
Here $\Omega$ is the solid angle, within which a point outside the
neutrino sphere can receive the neutrinos. It is calculated by the geometric calculations (see Figure \ref{neusp3dim}).
As stated, the neutrino emission from each point on the neutrino sphere
is assumed to be isotropic and take a Fermi distribution with a
vanishing chemical potential.
The Eq. (\ref{bef_final}) is expressed with the local neutrino flux,
$j_{\nu}$, which is in turn fixed from the neutrino luminosity obtained
in the simulations:
\begin{equation}
Q_{\nu, B = 0}^{+} = \frac{(3 c_{\rm A}^2 + 1)}{4\, \pi} \sigma_{0} n_{\rm n}
\frac{\displaystyle{\frac{F_5(0)}{F_3(0)}} (k T_{\nu})^2}{(m_e c^2)^2} \,\,j_{\nu}\,  \Omega,
\label{last}
\end{equation}
where we used the relations,
\begin{equation} 
\langle\epsilon^2_{\nu}\rangle = F_{5}(0)/F_3(0) (k T_{\nu})^2,
\label{zero}
\end{equation}
with $F_{5}(0)$ and $F_3(0)$ being the Fermi integrals and 
\begin{eqnarray}
j_{\nu} &=& c \,\,\frac{2 \pi}{(hc)^3} \int \!\!d \epsilon_{\nu}\, \epsilon^3_{\nu} \,f_{\nu}(\epsilon_{\nu}) \int_{0}^{1} d \mu\, \mu  \nonumber \\
&=& c \,\, \frac{\pi}{(hc)^3} \int \!\!d \epsilon_{\nu}\,\,\epsilon^3_{\nu} \,f_{\nu}(\epsilon_{\nu}).
\label{jneu}
\end{eqnarray}
By the same procedure as above, the heating rate arising from the parity
violating effects can be given as follows,
\begin{eqnarray}
\Delta Q^{+}_{\nu,~B \neq 0} & = &
  \frac{3 c_{\rm A}^2 +1 }{4 \pi}\sigma_{0}~n_{\rm n}\Biggl[
\Bigl(\int d\Omega~ {\mbox{\boldmath$\hat{n}$}} \cdot \mbox{\boldmath$\hat{\rm{B}}$}\Bigr)~\frac{1}{(m_e c^2)^2} \times \nonumber \\
& & \Biggl( \frac{1}{2}{\hbar c e B}\frac{c_{\rm V}^2 - c_{\rm A}^2}{{c_{\rm V}^2 + 3 c_{\rm A}^2}} -  2~\frac{c_{\rm A}(c_{\rm A} + c_{\rm V})}
{c_{\rm V}^2 + 3 c_{\rm A}^2}\frac{F_4{(0)}}{F_3{(0)}}~(\mu_{\rm B n}B)~ (k_{\rm B} T_{\nu})
 \nonumber\\
& & - 2~\frac{c_{\rm A}(c_{\rm A} - c_{\rm V})}
{c_{\rm V}^2 + 3 c_{\rm A}^2}~\frac{\mu_{\rm B p}B}{k_{\rm B} T}~\frac{F_5{(0)}}{F_3{(0)}}(k_B T_{\nu})^2  \nonumber\\ 
& & - 2~\frac{c_{\rm A}(c_{\rm A} - c_{\rm V})}
{c_{\rm V}^2 + 3 c_{\rm A}^2}~\frac{F_4{(0)}}{F_3{(0)}}~({\mu_{\rm B p}B})
~({k_{\rm B} T_{\nu}})
 \Biggr)j_{\nu} 
\Biggr].
\label{longformula}
\end{eqnarray}
Here $F_{4}(0)$ is the Fermi integral.
The inner product of $ {\mbox{\boldmath$\hat{n}$}} \cdot 
 \mbox{\boldmath$\hat{\rm{B}}$}$ can be obtained at each point by the
 geometry. 

Using the hydrodynamic quantities at the final time of the numerical 
simulations ($\sim 50$ msec) as a background,
we demonstrate how much the rotation-induced anisotropic neutrino
heating can be affected by the parity-violating corrections in the
following.

\begin{figure}
\begin{center}
\plotone{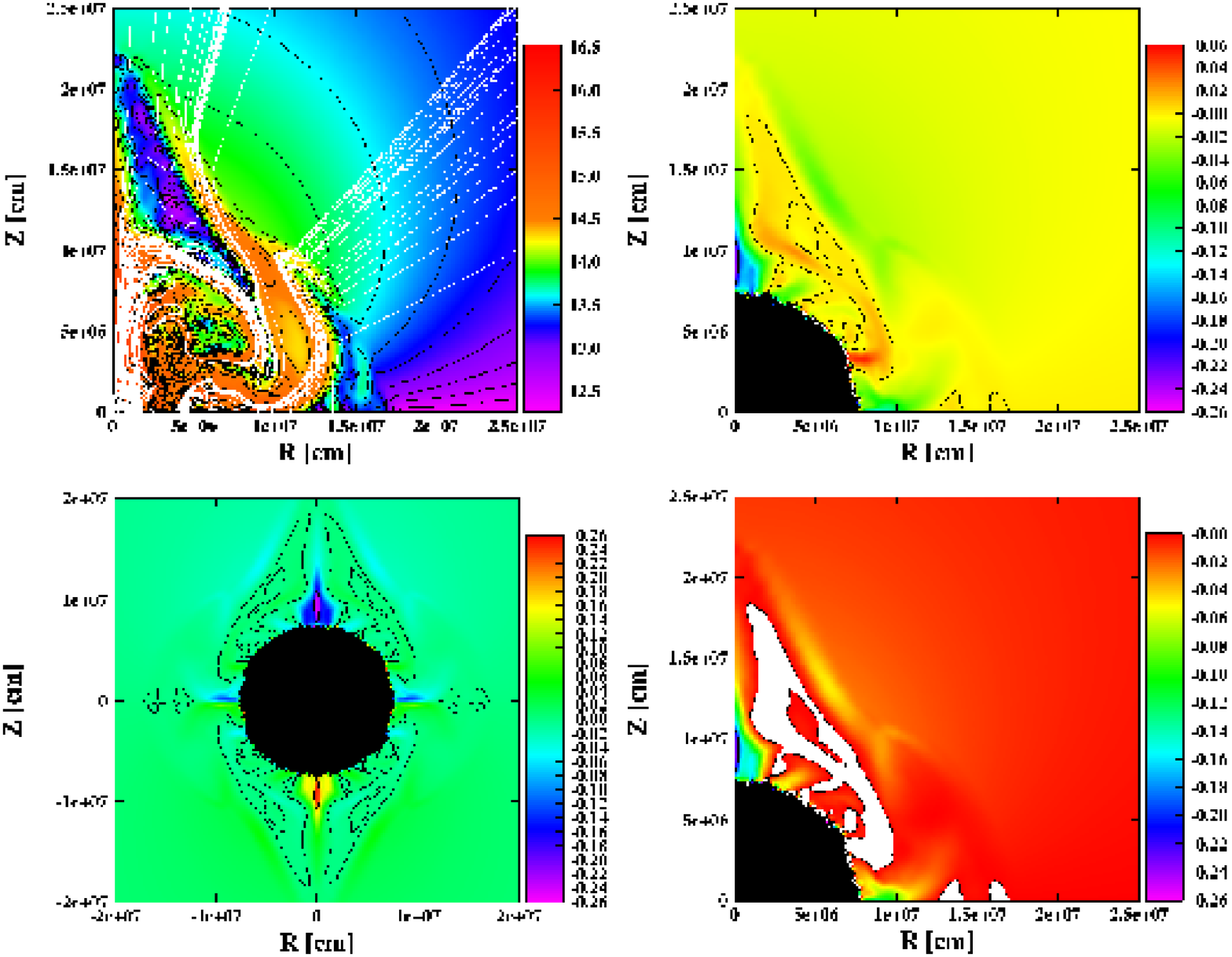}
\caption{Various quantities for the model SSU-2. Top left panel
 shows contour of the logarithm of the poloidal magnetic fields ($\log
 [B_{p}~{(\rm G})]$) with the magnetic field lines. Top right panel
 represents the ratio of the neutrino heating rate 
 corrected from the parity-violating effects, $\Delta Q_{\nu, \rm{B} \neq 0}^{+}$ to the heating rate without the corrections, $Q_{\nu, \rm{B} = 0}^{+}$. Note in the panel that the values of the color scale are expressed in percentage and that the central black region represents the region inside the neutrino sphere.
Bottom right panel is the same with the top right panel except
 that the bottom right panel only shows the regions with the negative
 values of the ratio. Thus the white region shows the regions with the
 positive values of the ratio. Bottom left panel shows the contour of
 the ratio in the whole region, which is prepared in order to see the
 global asymmetry of the heating induced by the strong magnetic fields.}
\label{example_hikaku}
\end{center}
\end{figure}

First of all, we take the model SS(U)-2 as an example model.
 The configuration of the poloidal magnetic fields at the final state 
is presented in the top left panel of Figure \ref{example_hikaku}.
It is seen from the panel that the poloidal magnetic fields are rather 
straight and parallel to the rotational axis in the regions near the
rotational axis, on the other hand, bent in a complex manner, in the other
central regions. This can be understood as follows.
The core bounce occurs anisotropically due to rotation.
This induces the convective motions. The magnetic fields are bent with the convective
motions because the magnetic fields are frozen 
in fluid elements. This is because the ideal MHD is assumed 
in our simulations. 
The top right panel of Figure \ref{example_hikaku} shows 
 the ratio of the neutrino heating rate 
 corrected from the parity-violating effects, $\Delta Q_{\nu, \rm{B} \neq 0}^{+}$ to the heating rate without the corrections, $Q_{\nu, \rm{B} = 0}^{+}$.  Note in the panel that the values of the color
scale are expressed in percentage and that the central black region
represents the region inside the neutrino sphere. It should be
 remembered that we
took the two choices of $T/|W|$ as the initial condition. 
In the models with the smaller $T/|W| = 10^{-2}~\%$, the shapes of the 
neutrino sphere are found to be rather spherical and the resultant
neutrino heating from the neutrino sphere becomes almost isotropic (see the
 models whose names end with ``-2'' in the right panel of Figure
 \ref{fig3} and Table \ref{heat_hikaku}).
In such models, it is found that
the initial rotation rate and magnetic field are not 
large enough to make the shapes of the
 neutrino sphere deviated from being spherically symmetric as in case of
the pure rotation \citep{kotake}.
In the top right panel, it is seen that the values of the ratio become
 negative in almost all the regions 
(see also the bottom right panel of Figure
 \ref{example_hikaku} for clarity). This means that the heating rate is reduced by
 the magnetic fields than that without. Especially, this tendency is most remarkable in the regions
 near the rotational axis and the surface of the neutrino sphere (see
 the regions colored by blue in the top right panel of Figure 
\ref{example_hikaku}). These features can be understood as follows.
Let us remember that the first term in Eq (\ref{longformula}) represents
the electron contribution arising from the ground landau level, and
the other represents the contribution from the neutron and proton
polarizations. By an order-of-magnitude estimation, it can be readily
found that the first term dominates over the other in the central core,
with typical values of $B \geq 10^{15}~{\rm G}~,T \sim O({\rm
 MeV})$. Since the coefficient 
of the first term, which is proportional to the coupling constants of weak
interactions ($\propto c_{\rm V}^2 - c_{\rm A}^2$, where $c_{\rm V} =
 1, c_{\rm A} = 1.26$) is always negative, then the suppression or the enhancement of the heating rate through
parity-violating effects is determined by the signs of the inner
product of $ {\mbox{\boldmath$\hat{n}$}} \cdot 
 \mbox{\boldmath$\hat{\rm{B}}$}$.
As mentioned above, the
magnetic fields are almost aligned and parallel to the rotational axis
 in the regions near the rotational axis. With the rather spherical
 neutrino emission, the values of the inner product becomes positive in 
the regions, which results in the reduction of the heating rate. On the other hand, the values of the ratio is found to become positive in some regions, where the poloidal magnetic fields are bent
due to convections (see the regions colored by red and colored by white 
in the top and bottom right panels of Figure \ref{example_hikaku}, respectively). In fact, the contributions of the inner product
are negative in such regions. In the bottom left panel of Figure 
\ref{example_hikaku}, the contour of the ratio in the 360 latitudinal
 degrees region of a star, is prepared in order to see the global asymmetry 
of the heating. 
Since our simulations assume the equatorial symmetry, the above features
 in the northern part of the star become reverse for the southern part. 
These features on the
 global asymmetry of the neutrino heating rates are almost common to the 
other models. 
In the following, we state some important features of the global
 asymmetry produced by the difference of the initial rotation rate 
or the initial configurations of the magnetic fields.

\begin{figure}
\begin{center}
\plotone{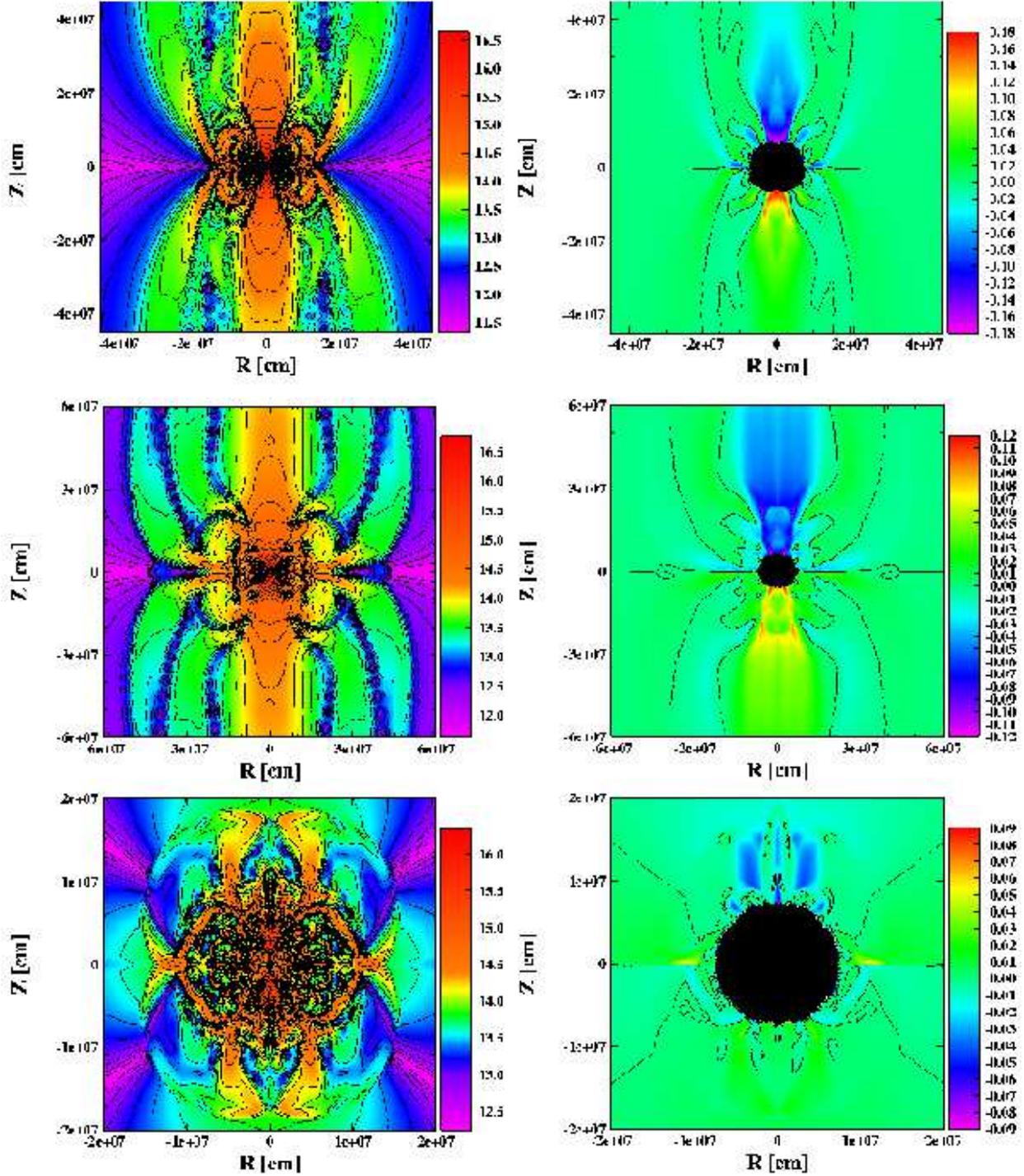}
\caption{The contour of the logarithm of the poloidal magnetic fields (G)
 (left panels) and the contour of $\Delta Q_{\nu, \rm{B} \neq 0}^{+}/Q_{\nu, 
{\rm B} = 0}^{+}$ in percentage (right panels). The top, middle, and 
bottom panels,
 correspond to the models CS(NU)-2, SS(NU)-1, and CS(Q)-2, respectively.}
\label{heat_mag}
\end{center}
\end{figure}

The configurations of the poloidal magnetic fields and the ratio of
$\Delta Q_{\nu, \rm{B} \neq 0}^{+}/Q_{\nu, \rm{B} = 0}^{+}$ for some
representative models at the final states are shown in Figure \ref{heat_mag}.
In the models, whose initial profile of the rotation law and the
magnetic field is cylindrical with high central concentration 
(the models whose name has ``CS(NU)``), the shapes of the shock wave
becomes very collimated near the rotational axis (see the bottom left
panel of Figure \ref{fig1}). The final
profile of the magnetic fields is also centrally concentrated with the 
strong magnetic fields ($B \sim 4 \times 10^{16}$ G), which is favorable
for the heating through parity-violating effects (see the top left and
right panels of Figure \ref{heat_mag}). As a result, the regions, where
the equatorial symmetry of the neutrino heating are broken, become
rather wider at the north and the south poles than the other models,
with the different initial profiles of the rotation and magnetic fields, 
while fixing the initial strength of rotation. If the initial strength of rotation is
more large, the neutrino sphere is deformed to be more oblate due to
the centrifugal force (see the black region of the middle right 
panel of Figure \ref{heat_mag}). Since the radii of the neutrino spheres
tend to be smaller near the rotational axis than that in the equatorial
plane (see Table \ref{heat_hikaku}), the solid angle within
which a point outside the neutrino sphere receives the rays from the
neutrino sphere is enlarged for the vicinity of the rotational axis.
This effect makes the neutrino heating near the rotational axis more 
efficient as in case of pure rotation \citep{kotake}. 
This rotation-induced anisotropic neutrino heating enhances the effect
of the symmetry of the heating induced by the parity-violating
corrections (see the values of $R_{\rm mag}$ in Table \ref{heat_hikaku} and
compare the values between the models with the same name except for the final
number, for example, SL(U)-2 and SL(U)-1).
Furthermore, the regions, which are heated asymmetrically, are 
enlarged for the faster rotating model than the slower rotating model 
(compare the top right and the middle right panel of Figure \ref{heat_mag}).

Finally, we refer to the model whose initial profile of the magnetic
field is quadrupole (see the bottom panels of Figure \ref{heat_mag}).  
In such models, it is found that  there appear the regions with
the strong magnetic fields, which seem like the horns, near the
rotational axis ranging from $\sim 100~{\rm km} < Z < 200~{\rm km}$ 
(see the bottom left panel of Figure \ref{heat_mag}).
As a result, the heating regions influenced by the
parity-violating corrections, are found to become rather off-axis (see
the bottom right panel of Figure \ref{heat_mag}).  

\section{Discussion \label{s4}}
In our computations, only electron neutrinos are treated
approximately. Since all the numerical simulations are limited to rather
early phases ($\sim$ 50 msec),
it may not be a bad approximation. However, this is a certain limitation to 
the study of the later phases, which are more important to investigate the 
outcome of the north-south neutrino heating asymmetry discussed here. 
As already mentioned, if we take into the cooling, no heating region is
found at these early times.  
As shown, for example, by \citet{tomp}, the heating region or (the gain
region) appears more than $50 \sim 100$ msec after core bounce. Hence,
our estimate of anisotropic neutrino heating is admittedly a bold
extrapolation to the later phases, which we have not computed
yet. 
As stated, the multidimensional MHD simulations 
which solve the neutrino transport with the parity-violating reactions
are required in order to compute the relation between the asymmetric 
neutrino heating induced by the parity-violating corrections
and the pulsar-kick.
Although we hope that the general features obtained in this study
will be taken over later on, further numerical investigations
should be required in order to check the validity of our results.

As for the initial configurations of the poloidal magnetic fields, we 
assume that they have the same profile as the rotation.
The assumption may not be unnatural because the angular
momentum transfer in quasistatic stellar evolution is tightly connected
to the formation of the magnetic fields \citep{parker,spruit}. However,  
multidimensional and self-consistent calculations are required to be
done for the understanding of 
the magnetorotational evolution of supernova progenitors.    

\section{Conclusion \label{s5}}
We performed a series of magnetohydrodynamic simulations of supernova
cores. As for the microphysics, we employ a realistic 
equation of state based on the relativistic mean field theory and take
 into account electron captures and the neutrino transport via the
 neutrino leakage scheme.
Since the profiles of the angular momentum and the magnetic fields of
strongly magnetized stars are still quite uncertain, 
we constructed 11 initial models changing the combination of the strength of
the angular momentum, the rotations law, the degree of differential
rotation, and the profiles of the magnetic fields in order to cover as
wide a parametric freedom as possible. 
Based on the numerical simulations, we estimated how the rotation-induced
anisotropic neutrino heating is changed by the parity-violating
corrections of neutrinos under the strong magnetic fields.
By so doing, we investigated the global asymmetry of the neutrino
heating in a strongly magnetized supernova core.
As a result, we found the following.

1. At the prompt-shock timescale ($\sim 50$ ms after core bounce), 
the toroidal magnetic fields are formed mainly by the wrapping of the initial
poloidal magnetic fields. The strength of the toroidal magnetic fields 
is smaller than the poloidal ones at the timescale. 
Although the growth timescale of MRI corresponding to the
fastest growing mode of the instability is within the prompt-shock
timescale, it is much longer than the growth of the wrapping timescale.
The inefficiency of the growth of MRI is due to the initial slow
rotation rates prior to core-collapse, which reconciles with the results 
of the state-of-the-art stellar evolution calculations.
 
2. Whether the parity-violating corrections in the strong magnetic
fields suppress or enhance the heating rate than that without the corrections
is found to be simply determined by the inner
product of the unit vector of incoming neutrinos ($ {\mbox{\boldmath$\hat{n}$}}$) and the unit vector
along the magnetic field direction ($\mbox{\boldmath$\hat{\rm{B}}$}$).
In the models except those with the quadrupole magnetic fields initially, the
poloidal magnetic fields near the rotational axis are almost aligned 
and parallel to the axis at the prompt-shock timescale. 
Although the neutrino emission is almost spherical in the slowly rotating
models with the initial values of $T/|W| = 10^{-2}~\%$, the values of
the inner product becomes positive in the regions by the geometry of the
magnetic field lines. This results in the reduction of the neutrino
heating rates. In the faster rotating models with the initial
values of $T/|W| = 10^{-1}~\%$, the rotation-induced anisotropic
neutrino heating is found to more enhance the asymmetric heating
induced by the parity-violating correction than in case of the slower
rotating models. This is because the rotation-induced neutrino heating
heats the regions near the rotational axis more efficiently, where the 
magnetic fields are typically strongest. The ratio of the correction to
the heating rate to the heating rate without the magnetic fields is
found to range from $0.08 \%$ to $0.42 \%$.

From the above results, 
the parity-violating corrections reduces about
$\lesssim 0.5 \%$ of the heating rate than that without the magnetic fields 
in the vicinity of the north pole, on the other hand, enhances about
$\lesssim 0.5 \%$ in the vicinity of the south pole of a star.
Based on the results, we speculate that 
the neutron star might be kicked toward the north pole in the subsequent time,
if the asymmetry of the neutrino heating in the north and the
south poles develop in the later phases. 
This is because \cite{scheck}.
 recently pointed out the random velocity
perturbation with an amplitude of $\sim 0.1$ \% added artificially at several
milliseconds after bounce can be a seed of the global asymmetry of
the supernova shock and a cause of the pulsar-kick.
In their models. the orientation of pulsar-kicks  
depends stochastically on the initial random perturbation. This is the 
case for the observed pulsars with the canonical magnetic fields ($B \sim 
10^{12}$ G).
However, the asymmetry of the neutrino heating 
in the strong magnetic fields ($B \geq 10^{15}$ G)  considered here might have influence on the alignment of spin and 
the space velocity of the pulsar associated with the formation of the magnetars. Although no pulsar kicks with the very strong magnetic 
fields, have been observed so far, the alignment of spin and the space 
velocity of the magnetars is expected to be observed by  
future observations. 

As stated earlier, our simulations are crude in the treatment of 
microphysics compared with recent spherical models. 
This might be justified as long as we study the earlier phase in the 
prompt-explosion timescale. We are currently developing a
two-dimensional neutrino transport code, which will be necessary to 
test the validity of implications obtained in this study 
 (Kotake et al. 2004 in preparation).

\acknowledgements{K.K is grateful to K. Sumiyoshi for providing us
with the tabulated equation of state, which can be handled
without difficulties and to M. Shimizu for supporting computer
environments. K.K also thanks H. Sawai and T. Takiwaki for
valuable discussions. The numerical calculations were partially done on the
supercomputers in RIKEN and KEK (KEK supercomputer Projects No.02-87 and
No.03-92). This work was partially supported by 
Grants-in-Aid for the Scientific Research from the Ministry of
Education, Science and Culture of Japan through No.S 14102004, No.
14079202, and No. 14740166. }

\clearpage



\clearpage

\clearpage

\clearpage


\clearpage




\clearpage






\end{document}